# Microwave Performance of all MOCVD-grown AlScN/GaN MIS-HEMTs on Semi-Insulating GaN Substrates

Can Cao[1,*], Vijay Gopal Thirupakuzi Vangipuram[1,*], Abdul Mukit[1], Motahareh Helli[2], Jinwoo Hwang[2], Hongping Zhao[1, 2,a] and Wu Lu[1, b]

[1]*Department of Electrical and Computer Engineering, The Ohio State University, Columbus OH, USA*
[2]*Department of Material Science and Engineering, The Ohio State University, Columbus OH, USA*

*Equal contribution
a) Email: zhao.2592@osu.edu
b) Corresponding Author Email: lu.173@osu.edu

**Abstract:** We report on the design, fabrication, and characterization of all MOCVD-grown long-gate AlScN/GaN metal-insulator-semiconductor high electron mobility transistors (MIS-HEMTs) on semi-insulating GaN substrates. Devices with a gate length of 1 μm and gate–drain spacing of 0.9 μm exhibit a maximum drain current density of 1 A/mm, an on/off current ratio of 2×10$^5$, and a three-terminal breakdown voltage of 63 V. The device has near-ideal subthreshold characteristics with a subthreshold swing of 63 mV/dec and a current dispersion as low as 7.8 % at 10 V due to the excellent interfacial quality with a trap density (D$_{it}$) of $2.11 \times 10^{11}$ cm$^{-2}eV^{-1}$ and the semi-insulating GaN substrate with a low threading dislocation density. Small-signal RF measurements reveal an f$_T$/f$_{max}$ of 25.8/51.1 GHz, while large-signal load-pull characterization at 10 GHz demonstrates an output power density of 4.04 W/mm with a power-added efficiency of 22.7%. In addition, a minimum noise figure below 2.5 dB was measured over a wide drain current range from 100 mA/mm to 700 mA/mm below 6 GHz. These results extend previous demonstrations of short-gate MOCVD-grown AlScN/GaN HEMTs to the long-gate, high-voltage regime, confirming the robustness of this material system for both high-frequency and high-power device applications with favorable microwave noise performance.

Aluminum scandium nitride (AlScN) has recently emerged as a highly attractive barrier layer for GaN high-electron-mobility transistors (HEMTs) owing to its strong spontaneous and piezoelectric polarization, which enable substantially higher two-dimensional electron gas (2DEG) densities compared with conventional AlGaN barrier [1,2]. Importantly, AlScN can be compositionally tuned to achieve closely lattice matching with GaN, thereby combining the benefit of reduced interfacial defects with strong spontaneous polarization-induced charge enhancement [3]. This unique balance allows AlScN/GaN heterostructures to sustain high carrier densities without severely compromising mobility, while also offering excellent thermal stability and mechanical robustness. As a result, ScAlN/GaN HEMTs are increasingly recognized as promising candidates for next-generation high-power and high-frequency electronics, spanning applications from 6G communications to space and quantum devices [4-6]. Recent state-of-the-art demonstrations have validated this potential. Scaled AlScN/GaN HEMTs with a gate length smaller than 100 nm grown by molecular beam epitaxy (MBE) have achieved sheet charge densities above $4 \times 10^{13}$ cm$^{-2}$ [7], transconductance exceeding 700 mS/mm [8], cut-off frequencies f$_T$ higher than 150 GHz with f$_{max}$ over 300 GHz

[9,10], and RF output power densities >5 W/mm with >40% power-added efficiency in the X–Ka band [8]. While high-quality AlScN/GaN interfaces have been demonstrated via MBE, a critical step in adopting AlScN within barrier layers for RF HEMTs is the mass-manufacturability. Metal-organic chemical vapor deposition (MOCVD) is the preferred epitaxial technique that has been utilized for GaN HEMTs production. Recent studies have confirmed that AlScN/GaN HEMTs can be realized by MOCVD with competitive DC and RF performance [11,12]. These developments underscore the importance of further evaluating MOCVD-grown AlScN/GaN devices under different device architectures and operating regimes. However, the Schottky gate architecture in the MOCVD-grown AlScN/GaN HEMTs suffers from relatively high off-state leakage currents, which limit the on/off current ratio. To address this, in this work a metal–insulator–semiconductor (MIS) gate structure is designed on AlScN/GaN heterojunctions on a semi-insulating (S.I.) GaN substrate with a low threading dislocation density to suppress leakage and sharpen the subthreshold slope. We use devices with 1-μm gates for assessment of material quality and interface electrostatics in the absence of strong short-channel effects. We show such MOCVD-grown AlScN/GaN MIS-HEMTs have low leakage current, near-ideal substhreshold swing (SS) (63 mV/dec), low RF current dispersion (7.8%), and promising RF small signal, large signal, and microwave noise performance, highlighting the potential of MOCVD-grown AlScN/GaN MIS-HEMTs for high-power applications. These results establish a robust foundation for future work on scaled MOCVD-grown AlScN/GaN MIS-HEMTs targeting high-frequency applications.

MOCVD growth of the full epitaxial structure was done on an ammonothermal, Mn-doped S.I. GaN substrate. To suppress the buffer leakage, here we chose the ammonothermal GaN substrates with a low dislocation density (~ $2\times10^5$ cm$^{-2}$) [14]. Prior growth, the substrate was cleaned in a 1:1 HCl:DI water volumetrically diluted solution for 5 minutes, followed by 20 minutes of soaking in a 1:3 $H_2O_2$:$H_2SO_4$ pirahna solution. To suppress the Si contaminant incorporation at the GaN growth interface, immediately prior growth, the substrate was dipped in a dilute HF solution for 5 minutes. MOCVD growth of the heterostructure started with ~100 nm of unintentionally doped (UID) GaN growth. This was followed by the AlN/AlScN barrier layer growth. Following AlScN growth, in-situ SiNx was deposited on the sample for an estimated thickness of ~ 2 nm. An estimated growth rate of ~ 1 nm/min was utilized for the total barrier layer growth (AlN and AlScN). A total combined growth time of 12 minutes was used for the barrier layers. This AlN/AlScN/SiN gate stack results in an equivalent oxide thickness (EOT) of 4.6 nm, enabling efficient gate control for scaled devices. A key roadblock in achieving AlScN films with compositional levels of Sc via MOCVD, is the lack of a suitable Sc precursor with sufficient vapor pressure. A customized MOCVD setup with high-temperature heating capability for the metal-organic precursor is required to achieve AlScN growth. In this work, (MCp)$_2$ScCl heated to 120°C was utilized as the Sc metal-organic precursor, while trimethylaluminum (TMAl) was utilized as the Al precursor. TMGa was utilized for GaN growth, and ammonia (NH$_3$) was utilized as the source for N species. Hydrogen was used as a carrier gas for all layers of growth while silane was used as the Si precursor during in-situ SiN$_x$ deposition. Details on the MOCVD growth of AlScN and development of AlScN/AlN/GaN heterostructrures can be found in previously published work. [15] To confirm the quality of the grown heterostructure, scanning transmission electron microscopy (STEM) of the sample was performed. A Thermo Fisher Scientific Thermis Z microscope at 200 kV was utilized for the high-annular dark field (HAADF) STEM imaging performed.

The schematic of the fabricated devices is shown in Fig. 1(a). All lithographic steps in the device fabrication were carried out using direct-write optical lithography. First, mesa isolation was performed by inductively coupled plasma – reactive ion etching (ICP-RIE) dry etching with a $BCl_3$-based process to a depth of 175 nm, ensuring complete etching into the semi-insulating GaN substrate. Prior to ohmic metallization, the in-situ $SiN_x$ cap was removed by a HF-based wet etching process. Alloyed Ti/Al/Ni/Au (20/120/30/100 nm) ohmic contacts were then formed by electron-beam evaporation, followed by rapid thermal annealing at 865 °C for 30 seconds. After that, Ni/Au gates were then deposited, followed by passivation using 200 nm SiN by plasma-enhanced chemical vapor deposition (PECVD). The gate length ($L_g$), gate-to-source spacing ($L_{gs}$), and gate-to-drain spacing ($L_{gd}$) of representative devices were measured using scanning electron microscopy (SEM) to be 1 μm, 0.5 μm and 0.9 μm, respectively. An optical micrograph of the actual fabricated device is shown in Fig. 1(c).

Ohmic contact resistance of 0.55 Ω·mm was extracted through the transmission line method (TLM). Improved ohmic performance on thick Al-rich barriers may be achieved through processes like recess etching and $n^+$ GaN regrowth [8,13]. Measurements of the Hall-bar test structures with integrated gate electrodes on in-situ SiN layer fabricated alongside the HEMTs yielded a sheet carrier density of $1.12 \times 10^{13}$ $cm^{-2}$, a mobility of 813 cm²/V·s, and a sheet resistance of 685 Ω/□. These results confirm the presence of a high-density 2DEG channel suitable for high-frequency operation.

Fig. 1(b) shows the cross-sectional STEM imaging of the AlScN/AlN/GaN heterostructure. A crystalline structure across the complete heterostructure's cross-section is evident for the sample. A clear interfacial contrast is present between the GaN channel and the AlScN/AlN barrier. However, no obvious contrast is present for the cross-sectional image between AlN and AlScN regions. It should be noted that since AlN has a practical critical thickness of ~6.5 nm on GaN [16], the presence of a barrier layer with a ~11 nm thickness without any obvious signs of strain relaxation indicates sufficient Sc incorporation has been achieved within the barrier layer to increase the critical thickness, further confirming the incorporation of compositional levels of Sc within the barrier layer. The presence of a clear, sharp contrast at the interface between the GaN and AlN layer evident from the cross-sectional imaging further bolsters support of the high-quality nature of the heterointerface. This directly correlates with the low current dispersion levels evidenced within the devices tested and presented within this study. Additional evidence that the barrier layer is fully strained to the GaN layer is bolstered by the asymmetric $10\bar{1}5$ reciprocal space mapping shown in Fig. S1 of the same heterostructure grown on a GaN-on-sapphire template sample. From Fig. S1, it is also evident that there is a gradual change in the layer's composition since there is no separate peak associated with the AlScN/AlN barrier; but rather a continuous line that extends from the substrate. This suggests that the Sc incorporation gradually increases along the grown thickness of the barrier layer. Based on thick AlScN layers grown under the same conditions in previous work, it is estimated that the maximum Sc composition of the barrier layer is up to ~13% [15]. Fig. 1 (d) shows TEM-EDX color mapping along the heterostructure's cross-section. Looking at the elemental color maps, compositional levels of Sc incorporation is evident close to the AlScN/$SiN_x$ interface. It should be noted that, as previously stated, there is likely a variation in the Sc incorporation along the heterostructure's barrier layer. This can result in lower levels of Sc incorporation within the heterostructure that may not be detectable by EDX measurements. Atomic force microscopy performed immediately after MOCVD growth of the MIS-HEMT structure is shown in Fig.

S2. The 5 μm × 5 μm scan area yielded an Ra RMS of 0.372 nm; highlighting the smooth surface morphology of the grown structure on the Mn-doped S.I. GaN substrate.

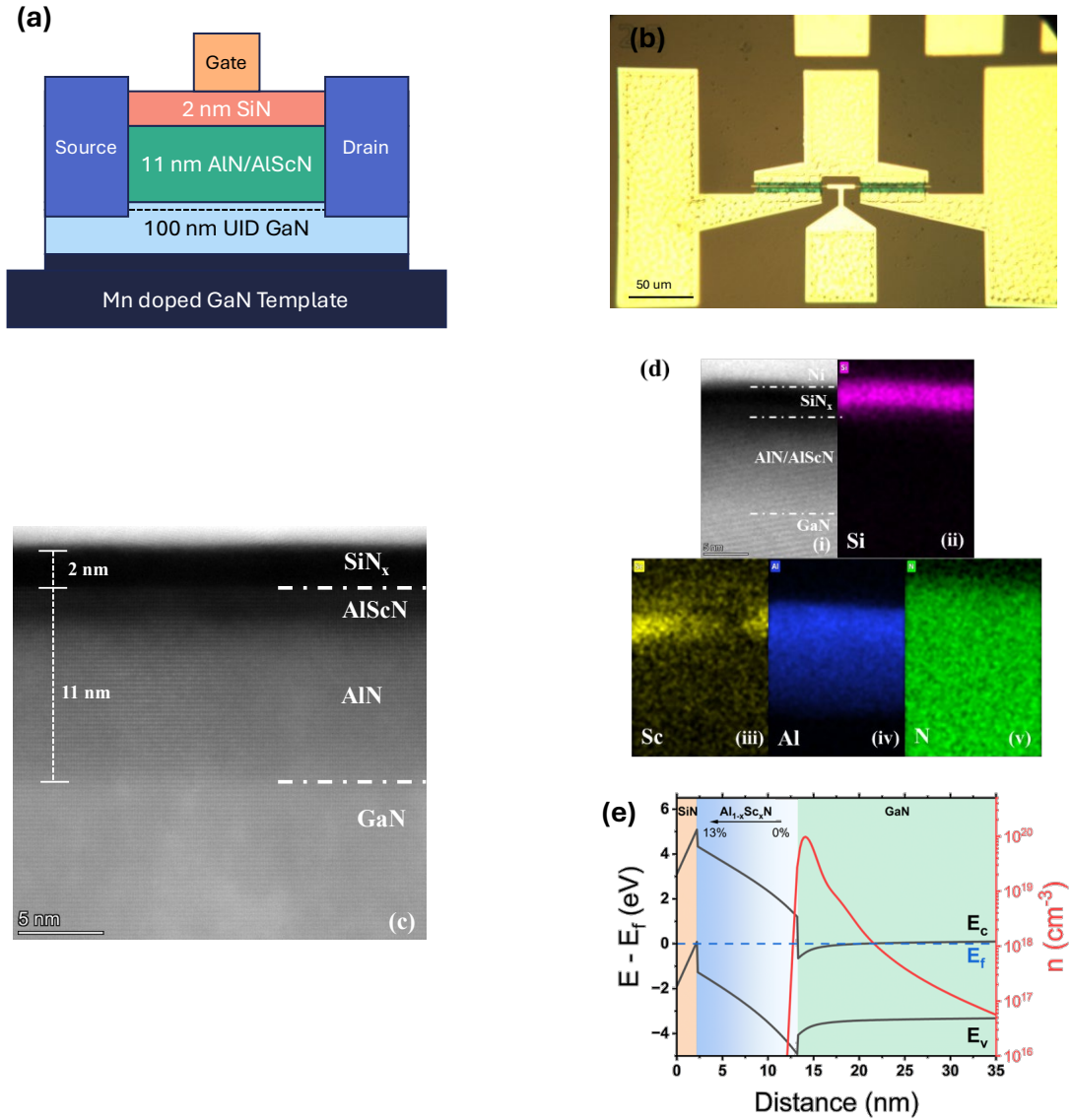

Figure 1. (a) Schematic of epitaxial and processed device structure of the AlScN/GaN MIS-HEMTs. (b) Microscope image of an AlScN/GaN MIS-HEMT. (c) High resolution STEM cross-sectional imaging of the AlScN/AlN/GaN MIS-HEMT structure. (d) STEM-EDX elemental color map of the MIS-HEMT layers, and (e) 1D simulated energy band diagram and electron density profile of the AlScN/GaN MIS-HEMT with SiN cap layer. The simulated 2DEG density is $2.09 \times 10^{13} cm^{-2}$.

To further connect the measured carrier density with the underlying polarization electrostatics, a 1D self-consistent band diagram simulation was performed using the experimentally observed graded AlScN composition profile and the barrier thickness determined from cross-sectional TEM. The simulated polarization-induced sheet charge density is $2.09 \times 10^{13} cm^{-2}$. It is noted that the presence of the SiN dielectric modifies the surface boundary condition and raises the conduction band in the barrier, leading to partial depletion of the polarization-induced charge at the AlScN/GaN interface. For comparison, the same AlScN/GaN heterostructure without the SiN cap layer has a 2DEG density of $3.39 \times 10^{13} cm^{-2}$ from simulation (Fig. S3). The difference between the measured value from Hall measurement and simulation is attributed to non-ideal effects not captured in the ideal electrostatic model such as interface states and partial compensation, which reduce the electrically active carrier density.

Figure 2(a) and (b) show DC transfer characteristics of a representative AlScN/GaN MIS-HEMT, measured at a drain voltage ($V_d$) of 10 V. The transconductance ($g_m$) peaks at 276 mS/mm at a gate voltage ($V_g$) of - 2.25 V. The device has a threshold voltage of - 3 V, determined from the gate bias with a channel current density of 1 mA/mm. The device demonstrated an on/off current ratio ($I_{on}/I_{off}$) of $2\times10^5$, which is one order higher than what have been reported ($< 10^4$) for MOCVD-grown AlScN/GaN HEMTs [11,12]. The device exhibited excellent gate control and pinch-off capabilities with a subthreshold swing (SS) of 63 mV/dec, approaching the Boltzmann theoretical limit at room temperature. We attribute this close-to-ideal SS to high quality epitaxial channel layer on the low threading dislocation density GaN substrate, excellent interfacial quality between the barriers and the channel, and effective electrostatic gate control. In fact, as shown in Fig. 2(b), the off-state current is limited by the gate leakage current at the level of $10^{-3}$ mA/mm, which is about two orders lower than reported MOCVD-grown AlScN/GaN HEMTs [11]. The low gate leakage and near-ideal SS of 63 mV/dec both point to well-controlled electrostatics and minimal trap-assisted leakage in the buffer, underscoring the potential of MOCVD-grown AlScN/GaN MIS-HEMTs for high-performance RF applications. The efficient gate control is achieved from the thin (4.6 nm) EOT of the AlN/AlScN/SiN gate stack. To provide deeper insight into the MIS-HEMT gate stack and interface quality, capacitance-voltage (C-V) measurements at different frequencies were performed. Figure 2(c) shows the C-V characteristics measured from 10 kHz to 2 MHz. The device exhibited essentially no frequency dispersion, which is consistent with the presence of interface states responding to the AC signal. The conductance-frequency (G-f) measurement was also performed, as shown in S4. The extracted interface trap density ($D_{it}$) is $2.11 \times 10^{11}$ cm$^{-2}eV^{-1}$. Consequently, the subthreshold swing (SS) can be calculated as $SS = 2.3\frac{kT}{q}\left(1+\frac{C_d+C_{it}}{C_i}\right) \approx 2.3\frac{kT}{q}\left(1+\frac{qD_{it}}{C_g}\right) \approx 63.2 \ mV/dec$, which is consistent with SS measured from DC measurements of HEMTs. This confirms that the low density of interface states enables the near-ideal SS.

Figure 2(d) illustrates three-terminal breakdown characteristics measured at $V_g$ = - 4 V. The breakdown voltage ($V_{br}$) is defined as the drain voltage corresponding to a leakage current of 1 mA/mm. Hard off-state breakdown occurred at 63 V, corresponding to an average gate to drain electric field ($E_{br}$) of 0.75 MV/cm. This result confirms the high-voltage capability of the long-gate device, consistent with its potential use in power RF applications. Figure 2(e) shows DC and pulsed family IV performance sweeping $V_g$ from 1 V to - 4 V with a step of - 1 V. Maximum drain current density $I_{d, max}$ was measured to be 0.99 A/mm at $V_g$ = 1 V. An on-resistance ($R_{on}$) of 2.6 Ω·mm was extracted. At higher drain biases, a gradual reduction in the output current was observed from the DC family IVs, due to the self-heating effect in the active region. Pulse measurements without heating effect with a pulse width of 200 ns and a 1 ms period at quiescent bias conditions of Vg = 0 V, Vd = 0 V show a maximum current density of 1.13 A/mm. Isotrapping condition with quiescent biases of $V_g$ = - 4 V and $V_d$ = 10 V, e.g., well below the threshold voltage at off-state, was also measured. The current dispersion at $V_d$ = 10 V is only 7.8 %, which is significantly lower than what has been reported on AlScN/GaN HEMT devices [8,11]. The low current dispersion highlights the reduced trapping effects in the present devices. This minimal current collapse indicates suppressed dynamic trapping effects, attributed to the combined effect of the in-situ SiN and PECVD SiN passivation.

(a)
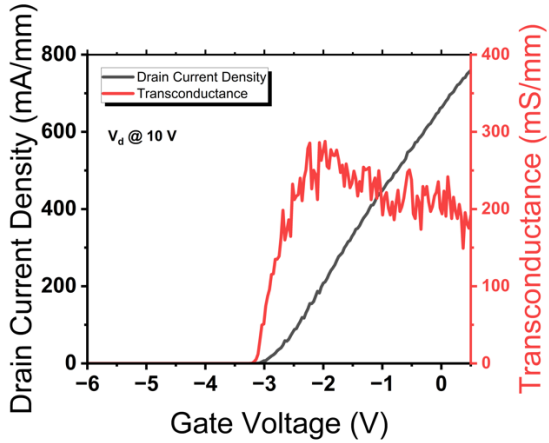

(b)
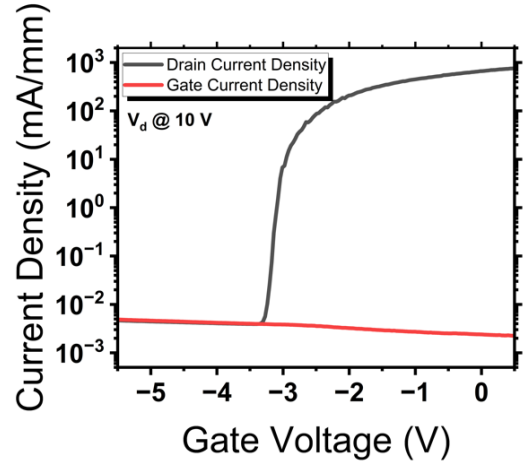

(c)
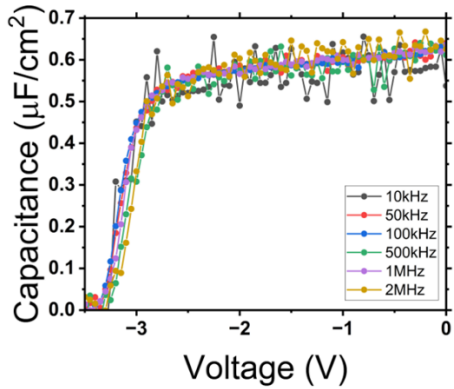

(d)
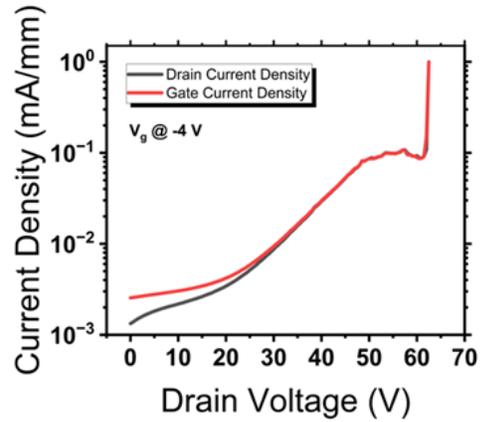

(e)
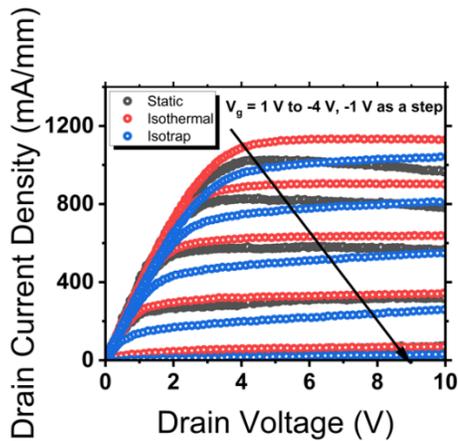

Figure 2. DC transfer characteristics in (a) linear and (b) semi-log scale of a representative AlScN/GaN MIS-HEMT, measured at $V_d$ = 10 V. (c) Capacitance-voltage characteristics of an AlScN/GaN gate diode measured at 100 kHz and 1MHz. (d) Three-terminal breakdown characteristics of a representative AlScN/GaN MIS-HEMT measured at $V_g$ = - 4 V. (e) DC and pulsed family IVs measured with a pulse width of 200 ns and a 1 ms period at quiescent bias conditions of isothermal: $V_g$ = 0 V, $V_d$ = 0 V and isotrap: $V_g$ = - 4 V, $V_d$ = 10 V. The gate was biased from 1 V to - 4 V in a step of - 1 V.

Small signal S-parameters were measured using an Agilent 8510C vector network analyzer from 1 to 45 GHz. The measured current gain ($|h21|^2$), maximum stable gain (MSG), and maximum available gain (MAG) are plotted in dB scale in Figure 3(a). The device exhibits a peak unity current gain cutoff frequency ($f_T$) of 25.8 GHz and a maximum oscillation frequency ($f_{max}$) of 51.1 GHz at $V_d$ = 10 V and $I_d$ = 261.9 mA/mm. These values demonstrate that, despite the relatively long gate length of 1 µm, the devices have competitive RF performance. Figure 3(b) summarizes the dependence of $f_T$ and $f_{max}$ on gate bias or the channel current. Both frequencies remain relatively stable and high in the gate voltage range of - 2.5 V to - 1.5 V (or $I_d$ = 32.7 mA/mm ~ 662 mA/mm), corresponding to the region around maximum transconductance ($g_{m, max}$), indicating robust channel control and high linearity. To further assess the potential of the MOCVD-grown AlScN/GaN platform for mm-wave applications, the effective electron velocity can be estimated by $v_{sat} \approx 2\pi f_T \times L_g = 1.62 \times 10^7 \, cm/s$, which is similar to state-of-the-art AlGaN/GaN HEMTs [19] and higher than previously reported AlScN/GaN HEMTs [9]. Assuming velocity preservation, scaling the gate length to 150 nm could theoretically yield an $f_T$ approaching 170 GHz, validating the intrinsic scalability of the material system for future Ka-band and sub-THz applications.

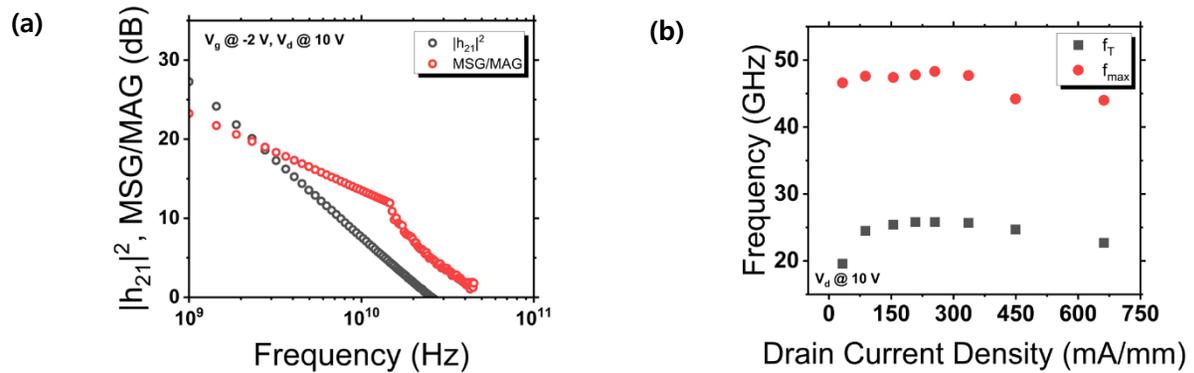

Figure 3. (a) Small signal characteristics of a representative AlScN/GaN HEMT, measured at $V_g$ = - 2 V and $V_d$ = 10 V. (b) Cut-off frequencies ($f_T$) and maximum oscillation frequencies ($f_{max}$) dependence of drain current density. The drain bias is 10 V. The gate is biased from – 2.5 V to – 1.5 V.

Continuous wave (cw) large signal measurements were performed at 10 GHz using a Focus Microwave load-pull system. Figure 4 shows the large-signal load-pull characteristics of the fabricated AlScN/GaN MIS-HEMTs measured at 10 GHz. Under Class-AB operation with load-pull for maximum output power, at a drain bias of $V_d$ = 30 V and a quiescent current of $I_d$ = 330 mA/mm, the devices delivered a maximum output power density ($P_{out}$) of 4.04 W/mm with a peak power-added efficiency (PAE) of 23%. The relatively low PAE is a result of self-heating effect under cw operation due to the low thermal conductivity of GaN substrate. The heating effect can be mitigated by substrate thinning and backside vias that are used at manufacturing of RF devices. Nevertheless, these results confirm that, even with a relatively long gate length of 1 μm, the devices can achieve competitive RF power performance. The present demonstration in the long-gate regime establishes a solid basis for subsequent short-gate development aimed at enhancing mm-Wave performance.

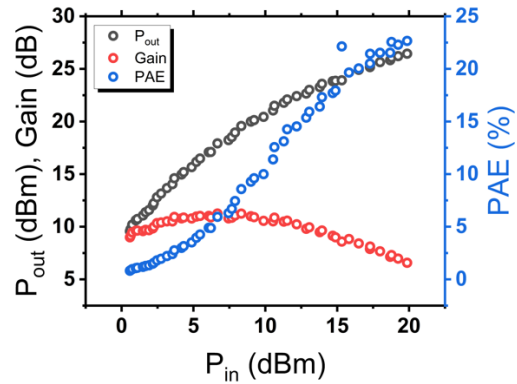

Figure 4. Large signal load-pull characteristics of a representative AlScN/GaN MIS-HEMT, measured at 10 GHz. The device was biased at $V_g$ = -1.3 V and $V_d$ = 30 V.

The microwave noise performance of the AlScN/GaN MIS-HEMTs was also measured using the cold-source measurement technique over the frequency range of 2-18 GHz [17]. Figure 5(a) shows the frequency dependence of the minimum noise figure ($NF_{min}$) and associated gain ($G_a$). During the measurement, the device was biased at the condition of $V_d$ = 10 V and $V_g$ = - 2.5 V ($I_d$ = 180 mA/mm). As shown, $NF_{min}$ reaches a minimum at a certain low frequency value and increases with frequency due to decrease in associated gain. The dependence of the $NF_{min}$ and $G_a$ on drain current density is shown in Fig. 5(b) with the drain bias fixed at 10 V while the gate voltage was varied to modulate the drain current. The frequency was set to the point of minimum noise figure determined in the frequency sweep. Overall, $NF_{min}$ increases with drain current, primarily due to increased diffusion noise. At high frequencies, the rise in noise figure is attributed to the reduction in associated gain because of the long gate length, which limits noise matching. Nevertheless, $NF_{min}$ values below 2.5 dB were maintained over a wide range of drain current density from 100 to 700 mA/mm. These values compare favorably with reported AlGaN/GaN HEMTs of similar gate lengths [18] and are consistent with the low subthreshold swing, low off-state and gate leakage current, and low current dispersion observed, all of which indicate suppressed trapping and good channel electrostatic control in the MOCVD-grown AlScN/GaN heterostructure on semi-insulating GaN substrates. To our knowledge, this is the first report on microwave noise performance of AlScN/GaN HEMTs. It should be pointed out that this device structure is highly scalable to shorter gate lengths for higher gain, larger bandwidth, and power density, as these 1 μm devices have a low threshold voltage. To achieve high $f_{max}$, it is well known that it requires not just gate length scaling but also the reduction of gate resistance ($R_g$) and output conductance ($g_{ds}$). The AlScN/GaN with a thin gate stack barrier or EOT enables a high aspect ratio even for heavily scaled devices, which is crucial for achieving high gain and suppressing short-channel effects at mm-wave frequencies.

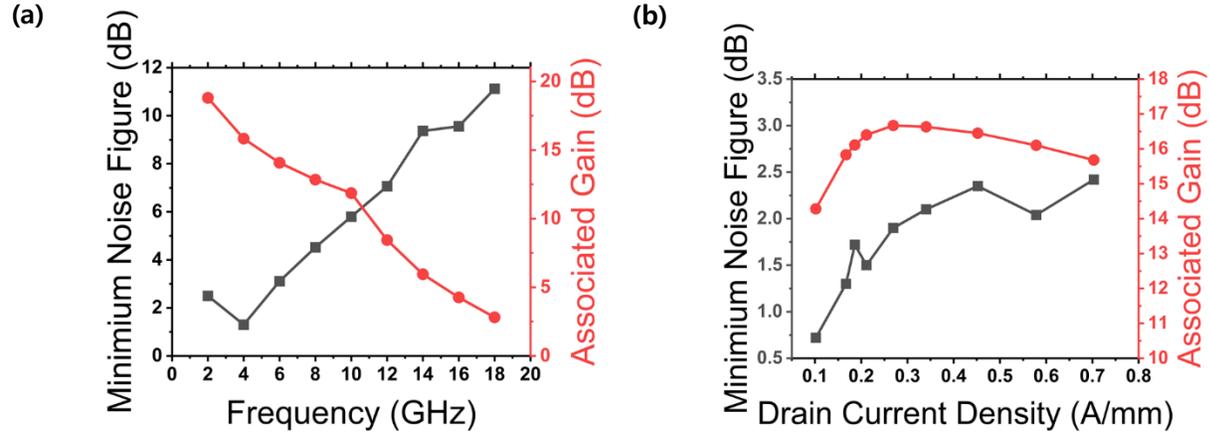

Figure 5. The minimum noise figure (black square) and associated gain (red circle) as a function of (a) frequency under the biasing condition of $V_g$ = - 2.5 V ($I_d$ = 0.18 A/mm) and (b) drain current density at 4 GHz ($V_g$ = - 2.8 ~ 0 V). The drain was biased at 10 V.

In summary, we have demonstrated AlScN/GaN HEMTs grown entirely by MOCVD on semi-insulating GaN substrates with competitive DC and RF performance. The devices achieved a maximum drain current of 0.99 A/mm, a peak transconductance of 276 mS/mm, and a breakdown voltage of 63 V. A high $I_{on}/I_{off}$ of $2\times10^5$, a close-to-ideal subthreshold swing of 63 mV/dec, and a low current dispersion of 7.8 % were observed, resulted from excellent electrostatics and suppressed trapping effects. Devices with a gate length of 1 μm exhibited an $f_T/f_{max}$ of 25.8/51.1 GHz, 4.04 W/mm output power density with 23 % PAE under Class-AB operation at 10 GHz. A minimum noise figure below 2.5 dB was also measured across a wide bias range at frequencies below 6 GHz. These complementary demonstrations establish a baseline for long-gate devices with low dispersion and near-ideal electrostatics, confirm the feasibility of MOCVD-grown AlScN/GaN for high-voltage operation and provide a foundation for future gate-length scaling and optimization. Further improvements in frequency performance are anticipated with reduced gate lengths, which will allow the intrinsic advantages of AlScN polarization to be fully leveraged.

**Supplementary Material**

Off-axis reciprocal space mapping (RSM) was performed along the asymmetric $10\bar{1}5$ reflection for the sample grown on a GaN-on-sapphire template. X-ray diffraction was performed on a Bruker D8 Discover utilizing a Cu K-α source. The sample's growth conditions for the GaN and AlScN/AlN barrier layers are the same as those used for the device sample. As seen in Fig. S1, the off-axis mapping reveals that the AlScN/AlN combinational barrier layer is fully strained within the in-plane direction to the GaN layer underneath. Atomic force microscopy (AFM) was performed on a Bruker Icon 3 AFM.

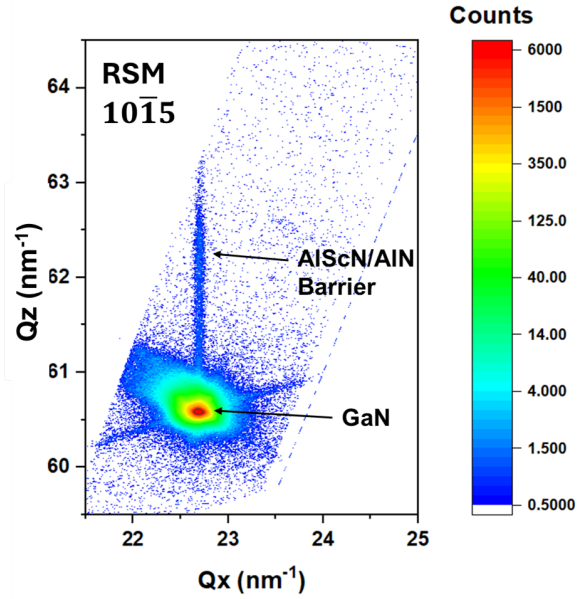

Fig. S1: Reciprocal space mapping of the asymmetric $10\bar{1}5$ reflection on a calibration sample on a GaN-on-sapphire template.

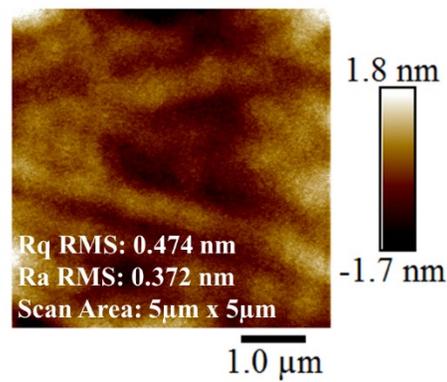

Fig. S2: AFM scan of the MOCVD-grown $SiN_x$/AlScN/AlN/GaN heterostructure on the Mn-doped GaN substrate.

1D self-consistent band diagram simulation was performed on AlScN/GaN HEMT structure without the SiN cap layer to study the impact of SiN layer on the sheet charge density. The simulated 2DEG density without SiN layer is $3.39 \times 10^{13} cm^{-2}$, significantly higher than the case with SiN layer.

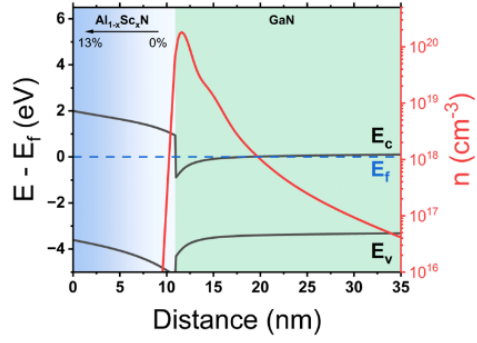

Fig. S3: 1D simulated energy band diagram and electron density profile of the AlScN/GaN MIS-HEMT without SiN cap layer. The 2DEG density is $3.39 \times 10^{13} cm^{-2}$ from simulation.

The conductance-frequency (G-f) measurement was also performed under a gate bias of $V_g = -3.5\ V$, as shown in Fig. S4. Using the AC-conductance method [S1], the interface trap density $D_{it}$ was extracted to be $2.11 \times 10^{11}\ cm^{-2} eV^{-1}$. Based on this value, the subthreshold swing (SS) can be calculated as $SS = 2.3 \frac{kT}{q}\left(1 + \frac{C_d + C_{it}}{C_i}\right) \approx 2.3 \frac{kT}{q}\left(1 + \frac{qD_{it}}{C_g}\right) \approx 63.2\ mV/dec$, which is observed in the DC measurements. The combination of the in-situ $SiN_x$ dielectric and the MIS architecture effectively suppresses gate leakage to $\sim 10^{-3}$ mA/mm enabling the steep switching slope.

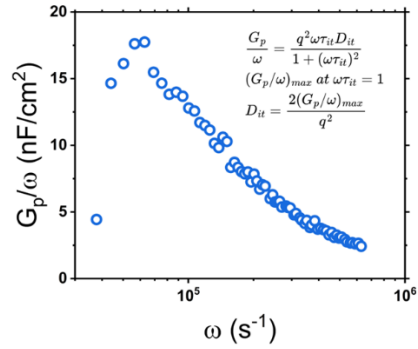

Fig. S4: Data for the conductance $G_p/\omega$ element of the AlScN/GaN HEMT, extracted from measured data under a gate bias of $V_g = -3.5\ V$.

**Acknowledgements**

This work was supported by the Army Research Office (Award No. W911NF-24-2-0210). Electron microscopy was performed at the Center for Electron Microscopy and Analysis (CEMAS) at The Ohio State University.

**Author Declarations**

**Conflict of Interest**

The authors have no conflicts to disclose.

**Data Availability**

The data that support the findings of this study are available from the corresponding author upon reasonable request.